\documentclass[]{article}
\usepackage{cmap} % fix search and cut-and-paste in Acrobat
\usepackage{ifthen}
\usepackage[T1]{fontenc}
\usepackage[utf8]{inputenc}
\usepackage{color}

\usepackage[font={small,it},labelfont=bf]{caption}
\usepackage{float}

\usepackage{arxiv}
\usepackage{scipy}
\makeatletter
\def\PY@reset{\let\PY@it=\relax \let\PY@bf=\relax%
    \let\PY@ul=\relax \let\PY@tc=\relax%
    \let\PY@bc=\relax \let\PY@ff=\relax}
\def\PY@tok#1{\csname PY@tok@#1\endcsname}
\def\PY@toks#1+{\ifx\relax#1\empty\else%
    \PY@tok{#1}\expandafter\PY@toks\fi}
\def\PY@do#1{\PY@bc{\PY@tc{\PY@ul{%
    \PY@it{\PY@bf{\PY@ff{#1}}}}}}}
\def\PY#1#2{\PY@reset\PY@toks#1+\relax+\PY@do{#2}}

\expandafter\def\csname PY@tok@w\endcsname{\def\PY@tc##1{\textcolor[rgb]{0.73,0.73,0.73}{##1}}}
\expandafter\def\csname PY@tok@c\endcsname{\let\PY@it=\textit\def\PY@tc##1{\textcolor[rgb]{0.25,0.50,0.56}{##1}}}
\expandafter\def\csname PY@tok@cp\endcsname{\def\PY@tc##1{\textcolor[rgb]{0.00,0.44,0.13}{##1}}}
\expandafter\def\csname PY@tok@cs\endcsname{\def\PY@tc##1{\textcolor[rgb]{0.25,0.50,0.56}{##1}}\def\PY@bc##1{\setlength{\fboxsep}{0pt}\colorbox[rgb]{1.00,0.94,0.94}{\strut ##1}}}
\expandafter\def\csname PY@tok@k\endcsname{\let\PY@bf=\textbf\def\PY@tc##1{\textcolor[rgb]{0.00,0.44,0.13}{##1}}}
\expandafter\def\csname PY@tok@kp\endcsname{\def\PY@tc##1{\textcolor[rgb]{0.00,0.44,0.13}{##1}}}
\expandafter\def\csname PY@tok@kt\endcsname{\def\PY@tc##1{\textcolor[rgb]{0.56,0.13,0.00}{##1}}}
\expandafter\def\csname PY@tok@o\endcsname{\def\PY@tc##1{\textcolor[rgb]{0.40,0.40,0.40}{##1}}}
\expandafter\def\csname PY@tok@ow\endcsname{\let\PY@bf=\textbf\def\PY@tc##1{\textcolor[rgb]{0.00,0.44,0.13}{##1}}}
\expandafter\def\csname PY@tok@nb\endcsname{\def\PY@tc##1{\textcolor[rgb]{0.00,0.44,0.13}{##1}}}
\expandafter\def\csname PY@tok@nf\endcsname{\def\PY@tc##1{\textcolor[rgb]{0.02,0.16,0.49}{##1}}}
\expandafter\def\csname PY@tok@nc\endcsname{\let\PY@bf=\textbf\def\PY@tc##1{\textcolor[rgb]{0.05,0.52,0.71}{##1}}}
\expandafter\def\csname PY@tok@nn\endcsname{\let\PY@bf=\textbf\def\PY@tc##1{\textcolor[rgb]{0.05,0.52,0.71}{##1}}}
\expandafter\def\csname PY@tok@ne\endcsname{\def\PY@tc##1{\textcolor[rgb]{0.00,0.44,0.13}{##1}}}
\expandafter\def\csname PY@tok@nv\endcsname{\def\PY@tc##1{\textcolor[rgb]{0.73,0.38,0.84}{##1}}}
\expandafter\def\csname PY@tok@no\endcsname{\def\PY@tc##1{\textcolor[rgb]{0.38,0.68,0.84}{##1}}}
\expandafter\def\csname PY@tok@nl\endcsname{\let\PY@bf=\textbf\def\PY@tc##1{\textcolor[rgb]{0.00,0.13,0.44}{##1}}}
\expandafter\def\csname PY@tok@ni\endcsname{\let\PY@bf=\textbf\def\PY@tc##1{\textcolor[rgb]{0.84,0.33,0.22}{##1}}}
\expandafter\def\csname PY@tok@na\endcsname{\def\PY@tc##1{\textcolor[rgb]{0.25,0.44,0.63}{##1}}}
\expandafter\def\csname PY@tok@nt\endcsname{\let\PY@bf=\textbf\def\PY@tc##1{\textcolor[rgb]{0.02,0.16,0.45}{##1}}}
\expandafter\def\csname PY@tok@nd\endcsname{\let\PY@bf=\textbf\def\PY@tc##1{\textcolor[rgb]{0.33,0.33,0.33}{##1}}}
\expandafter\def\csname PY@tok@s\endcsname{\def\PY@tc##1{\textcolor[rgb]{0.25,0.44,0.63}{##1}}}
\expandafter\def\csname PY@tok@sd\endcsname{\let\PY@it=\textit\def\PY@tc##1{\textcolor[rgb]{0.25,0.44,0.63}{##1}}}
\expandafter\def\csname PY@tok@si\endcsname{\let\PY@it=\textit\def\PY@tc##1{\textcolor[rgb]{0.44,0.63,0.82}{##1}}}
\expandafter\def\csname PY@tok@se\endcsname{\let\PY@bf=\textbf\def\PY@tc##1{\textcolor[rgb]{0.25,0.44,0.63}{##1}}}
\expandafter\def\csname PY@tok@sr\endcsname{\def\PY@tc##1{\textcolor[rgb]{0.14,0.33,0.53}{##1}}}
\expandafter\def\csname PY@tok@ss\endcsname{\def\PY@tc##1{\textcolor[rgb]{0.32,0.47,0.09}{##1}}}
\expandafter\def\csname PY@tok@sx\endcsname{\def\PY@tc##1{\textcolor[rgb]{0.78,0.36,0.04}{##1}}}
\expandafter\def\csname PY@tok@m\endcsname{\def\PY@tc##1{\textcolor[rgb]{0.13,0.50,0.31}{##1}}}
\expandafter\def\csname PY@tok@gh\endcsname{\let\PY@bf=\textbf\def\PY@tc##1{\textcolor[rgb]{0.00,0.00,0.50}{##1}}}
\expandafter\def\csname PY@tok@gu\endcsname{\let\PY@bf=\textbf\def\PY@tc##1{\textcolor[rgb]{0.50,0.00,0.50}{##1}}}
\expandafter\def\csname PY@tok@gd\endcsname{\def\PY@tc##1{\textcolor[rgb]{0.63,0.00,0.00}{##1}}}
\expandafter\def\csname PY@tok@gi\endcsname{\def\PY@tc##1{\textcolor[rgb]{0.00,0.63,0.00}{##1}}}
\expandafter\def\csname PY@tok@gr\endcsname{\def\PY@tc##1{\textcolor[rgb]{1.00,0.00,0.00}{##1}}}
\expandafter\def\csname PY@tok@ge\endcsname{\let\PY@it=\textit}
\expandafter\def\csname PY@tok@gs\endcsname{\let\PY@bf=\textbf}
\expandafter\def\csname PY@tok@gp\endcsname{\let\PY@bf=\textbf\def\PY@tc##1{\textcolor[rgb]{0.78,0.36,0.04}{##1}}}
\expandafter\def\csname PY@tok@go\endcsname{\def\PY@tc##1{\textcolor[rgb]{0.20,0.20,0.20}{##1}}}
\expandafter\def\csname PY@tok@gt\endcsname{\def\PY@tc##1{\textcolor[rgb]{0.00,0.27,0.87}{##1}}}
\expandafter\def\csname PY@tok@err\endcsname{\def\PY@bc##1{\setlength{\fboxsep}{0pt}\fcolorbox[rgb]{1.00,0.00,0.00}{1,1,1}{\strut ##1}}}
\expandafter\def\csname PY@tok@kc\endcsname{\let\PY@bf=\textbf\def\PY@tc##1{\textcolor[rgb]{0.00,0.44,0.13}{##1}}}
\expandafter\def\csname PY@tok@kd\endcsname{\let\PY@bf=\textbf\def\PY@tc##1{\textcolor[rgb]{0.00,0.44,0.13}{##1}}}
\expandafter\def\csname PY@tok@kn\endcsname{\let\PY@bf=\textbf\def\PY@tc##1{\textcolor[rgb]{0.00,0.44,0.13}{##1}}}
\expandafter\def\csname PY@tok@kr\endcsname{\let\PY@bf=\textbf\def\PY@tc##1{\textcolor[rgb]{0.00,0.44,0.13}{##1}}}
\expandafter\def\csname PY@tok@bp\endcsname{\def\PY@tc##1{\textcolor[rgb]{0.00,0.44,0.13}{##1}}}
\expandafter\def\csname PY@tok@fm\endcsname{\def\PY@tc##1{\textcolor[rgb]{0.02,0.16,0.49}{##1}}}
\expandafter\def\csname PY@tok@vc\endcsname{\def\PY@tc##1{\textcolor[rgb]{0.73,0.38,0.84}{##1}}}
\expandafter\def\csname PY@tok@vg\endcsname{\def\PY@tc##1{\textcolor[rgb]{0.73,0.38,0.84}{##1}}}
\expandafter\def\csname PY@tok@vi\endcsname{\def\PY@tc##1{\textcolor[rgb]{0.73,0.38,0.84}{##1}}}
\expandafter\def\csname PY@tok@vm\endcsname{\def\PY@tc##1{\textcolor[rgb]{0.73,0.38,0.84}{##1}}}
\expandafter\def\csname PY@tok@sa\endcsname{\def\PY@tc##1{\textcolor[rgb]{0.25,0.44,0.63}{##1}}}
\expandafter\def\csname PY@tok@sb\endcsname{\def\PY@tc##1{\textcolor[rgb]{0.25,0.44,0.63}{##1}}}
\expandafter\def\csname PY@tok@sc\endcsname{\def\PY@tc##1{\textcolor[rgb]{0.25,0.44,0.63}{##1}}}
\expandafter\def\csname PY@tok@dl\endcsname{\def\PY@tc##1{\textcolor[rgb]{0.25,0.44,0.63}{##1}}}
\expandafter\def\csname PY@tok@s2\endcsname{\def\PY@tc##1{\textcolor[rgb]{0.25,0.44,0.63}{##1}}}
\expandafter\def\csname PY@tok@sh\endcsname{\def\PY@tc##1{\textcolor[rgb]{0.25,0.44,0.63}{##1}}}
\expandafter\def\csname PY@tok@s1\endcsname{\def\PY@tc##1{\textcolor[rgb]{0.25,0.44,0.63}{##1}}}
\expandafter\def\csname PY@tok@mb\endcsname{\def\PY@tc##1{\textcolor[rgb]{0.13,0.50,0.31}{##1}}}
\expandafter\def\csname PY@tok@mf\endcsname{\def\PY@tc##1{\textcolor[rgb]{0.13,0.50,0.31}{##1}}}
\expandafter\def\csname PY@tok@mh\endcsname{\def\PY@tc##1{\textcolor[rgb]{0.13,0.50,0.31}{##1}}}
\expandafter\def\csname PY@tok@mi\endcsname{\def\PY@tc##1{\textcolor[rgb]{0.13,0.50,0.31}{##1}}}
\expandafter\def\csname PY@tok@il\endcsname{\def\PY@tc##1{\textcolor[rgb]{0.13,0.50,0.31}{##1}}}
\expandafter\def\csname PY@tok@mo\endcsname{\def\PY@tc##1{\textcolor[rgb]{0.13,0.50,0.31}{##1}}}
\expandafter\def\csname PY@tok@ch\endcsname{\let\PY@it=\textit\def\PY@tc##1{\textcolor[rgb]{0.25,0.50,0.56}{##1}}}
\expandafter\def\csname PY@tok@cm\endcsname{\let\PY@it=\textit\def\PY@tc##1{\textcolor[rgb]{0.25,0.50,0.56}{##1}}}
\expandafter\def\csname PY@tok@cpf\endcsname{\let\PY@it=\textit\def\PY@tc##1{\textcolor[rgb]{0.25,0.50,0.56}{##1}}}
\expandafter\def\csname PY@tok@c1\endcsname{\let\PY@it=\textit\def\PY@tc##1{\textcolor[rgb]{0.25,0.50,0.56}{##1}}}

\def\PYZbs{\char`\\}
\def\PYZus{\char`\_}
\def\PYZob{\char`\{}
\def\PYZcb{\char`\}}

\def\PYZsh{\char`\#}
\def\PYZpc{\char`\%}
\def\PYZdl{\char`\$}
\def\PYZhy{\char`\-}
\def\PYZsq{\char`\'}
\def\PYZdq{\char`\"}

\makeatother

\providecommand*{\DUrole}[2]{%
  \ifcsname docutilsrole#1\endcsname%
    \csname docutilsrole#1\endcsname{#2}%
  \else
    \csname DUrole#1\endcsname{#2}%
  \fi%
}

\providecommand*{\DUroletitlereference}[1]{\textsl{#1}}

\ifthenelse{\isundefined{\hypersetup}}{
  \usepackage[colorlinks=true,linkcolor=blue,urlcolor=blue]{hyperref}
  \usepackage{bookmark}
  \urlstyle{same} % normal text font (alternatives: tt, rm, sf)
}{}

\begin{document}
\title{Securing Your Collaborative Jupyter Notebooks in the Cloud using Container and Load Balancing Services}\author{Haw-minn Lu\\
Gary and Mary West Health Institute\\
La Jolla, CA 92037\\
\texttt{hlu@westhealth.org}\\
\And Adrian Kwong\\
Gary and Mary West Health Institute\\
La Jolla, CA 92037\\
\texttt{akwong@westhealth.org}\\
}\maketitle
\InputIfFileExists{page_numbers.tex}{}{}
\newcommand*{\docutilsroleref}{\ref}
\newcommand*{\docutilsrolelabel}{\label}
\newcommand*\DUrolecode[1]{#1}
\providecommand*\DUrolecite[1]{\cite{#1}}
\begin{abstract}Jupyter has become the go-to platform for developing data applications but data
and security concerns, especially when dealing with healthcare, have become
paramount for many institutions and applications dealing with sensitive
information. How then can we continue to enjoy the data analysis and machine
learning opportunities provided by Jupyter and the Python ecosystem while
guaranteeing auditable compliance with security and privacy concerns?  We will
describe the architecture and implementation of a cloud based platform based on
Jupyter that integrates with Amazon Web Services (AWS) and uses containerized
services without exposing the platform to the vulnerabilities present in
Kubernetes and JupyterHub. This architecture addresses the HIPAA requirements
to ensure both security and privacy of data. The architecture uses an AWS
service to provide JSON Web Tokens (JWT) for authentication as well as network
control. Furthermore, our architecture enables secure collaboration and sharing
of Jupyter notebooks. Even though our platform is focused on Jupyter notebooks
and JupyterLab, it also supports R-Studio and bespoke applications that share
the same authentication mechanisms. Further, the platform can be extended to
other cloud services other than AWS.\end{abstract}\keywords{data science, infrastructure, jupyter, rstudio}data science, infrastructure, jupyter, rstudio

\subsection{Introduction%
  \label{introduction}%
}

This paper focuses on secure implementation of Jupyter Notebooks and Jupyter
Labs in a cloud based platform and more specifically on Amazon Web Services
(AWS) though many architectures and methods described here are applicable to
other cloud platforms.

Project Jupyter includes Jupyter Hub, which provides a proxy and container
management. In particular the Zero to Jupyter Hub with Kubernetes project,
\DUrole{cite}{z2k8s} provides a framework to implement Jupyter Hub on a Kubernetes
platform but with many significant security drawbacks. Kubernetes is
notoriously difficult to secure and has many vulnerabilities that are not
addressed by default.

Security is paramount for applications that process sensitive data in areas
such as defense, finance, and healthcare. Broadly speaking, security
regulations can be characterized in terms of authentication (verifying the
credentials of users and their access to resources), encryption (data is
encrypted at rest and in transit), auditing (providing surveillance of key
resources) and vulnerability mitigation (antivirus and security updates).

In the architecture section, we describe how our architecture using AWS Elastic
Container Service (ECS) facilitates encryption at-rest and in-transit  and
integrates an application load balancer (ALB) for authentication.

In the Applications and Authentication section, we dive into the
details of the ALB and how JSON Web Tokens (JWT) facilitate integration with
Jupyter and RStudio.

In the Security and Compliance section, we address the encryption of the
underlying cloud architecture, auditing capabilities, and mitigation of
vulnerabilities.

Our specific implementation satisfies privacy and security concerns and can
serve as a starting point to develop customized solutions for related use
cases.

\subsection{Architecture%
  \label{architecture}%
}

There are two distinct levels of architecture described. The cloud architecture
comprises the various cloud services which is the lower layer of
virtualization. The container architecture is the top layer virtualization
built on top of the cloud architecture.

\subsubsection{Cloud Architecture%
  \label{cloud-architecture}%
}
\begin{figure}[]\noindent\makebox[\columnwidth][c]{\includegraphics[width=\columnwidth]{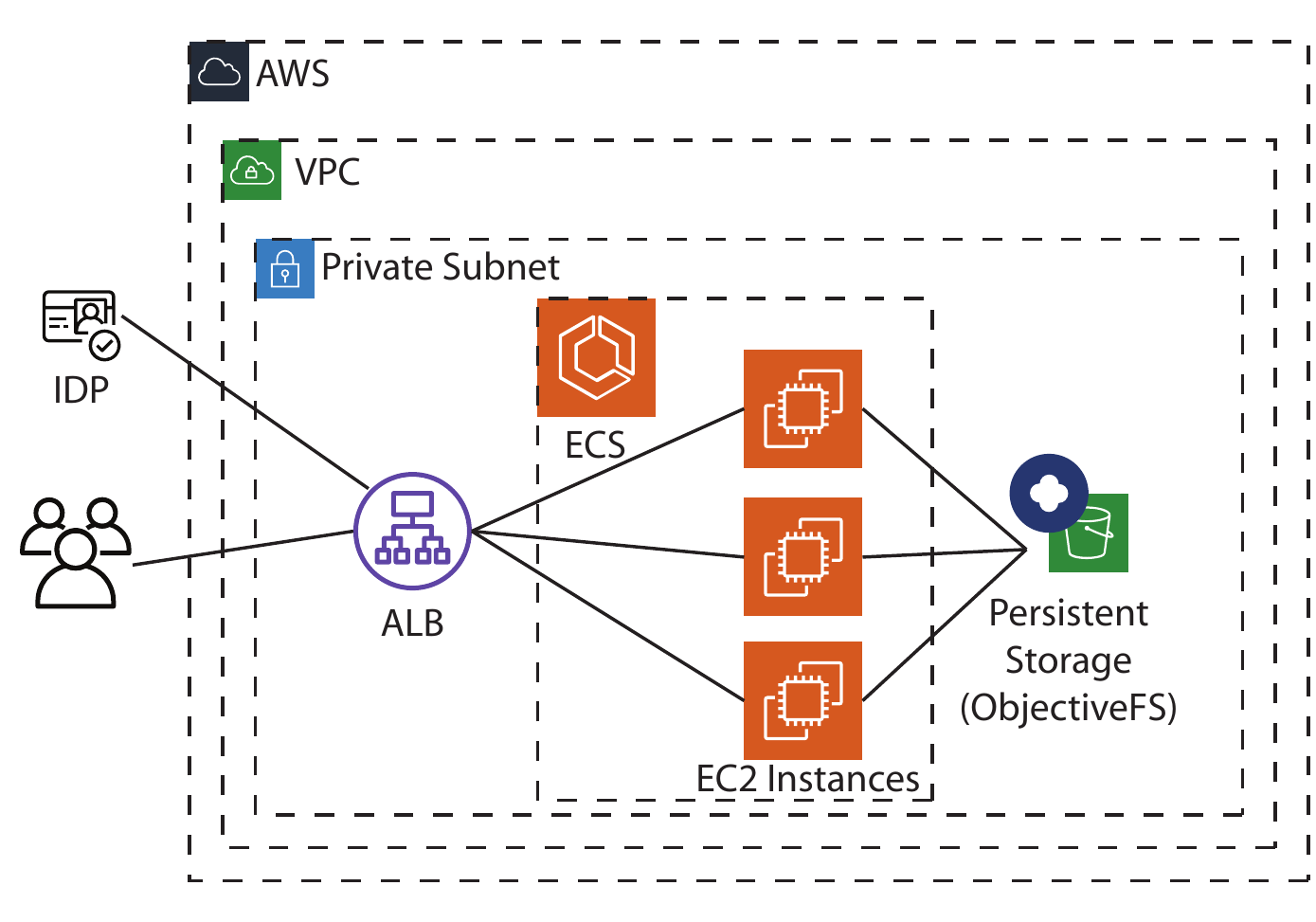}}
\caption{Cloud Architecture. \DUrole{label}{cloudfig}}
\end{figure}

The basic cloud architecture is shown in Fig. \DUrole{ref}{cloudfig}. It consists of
an identity provider (IDP) used to authenticate a user, an application load
balancer (ALB) to regulate user access through authentication, and a fleet of
elastic cloud computer compute (EC2) instances to instantiate the containers.
Finally, ECS manages the containers deployed on the EC2 cluster.

\paragraph{\textbf{Elastic Container Service}%
  \label{elastic-container-service}%
}

ECS is a container orchestration service. A container is instantiated as an ECS \emph{task}. ECS provides a resource called a \emph{task definition} that allow for the configuration of the container image, the environment variables, command override and container port.

Taking the most naive approach, ECS can be instructed to start a task based on a task definition. After the task has fully started, the host among the EC2 instances and the mapped port (the port on the EC2 node which is mapped to the container port) is known. At this point, one could write a monitoring function to detect when a task has started, retrieve the specific host and mapped port and create a listener rule for the application load balancer.

Instead of this cumbersome procedure, ECS provides another resource called a
\emph{service}. A service can manage many aspects of tasks within ECS including the
number of tasks and a \emph{target group} associated with the service. For our
purposes managing the number means selecting a desired count of 1 or 0
depending on whether the container is running or has been automatically culled due to inactivity. A
target group is a collection of host ports or serverless
AWS Lambda functions, to which a listener rule can direct network traffic. In
short by specifying a target group to a service, the host and mapped port are
automatically assigned to the target group when a task has fully started.

\paragraph{\textbf{Application Load Balancer}%
  \label{application-load-balancer}%
}

AWS's ALB can comprise multiple listeners to support multiple protocols. To
maintain security, enforcement of HTTPS should be maintained either by not
including a listener for HTTP or providing an HTTP listener that redirects all
requests to HTTPS.

AWS's ALB, through a listener, is able to direct external HTTPS requests to
various components. Based on listener rules, a request can be directed on the
basis of both the hostname and the path. As an example, we use a path to
specify a user and a service such as Jupyter (for example,
\texttt{\DUrole{code}{domain.com/user\_id/Jupyter}} or \texttt{\DUrole{code}{domain.com/user\_id/rstudio}}), this
allows us to give each user their own container.

Each listener rule maps a path, hostname or both to a particular target group.
Since we use an ECS service, we can assign a particular service to a target
group. The service then manages which ports and EC2 instances are part of the
target group.

While the ALB can enforce encryption from the end user to the ALB, the
container application (e.g., Jupyter) should also be configured to listen only
for HTTPS. In this manner, the communication from the end user to the ALB is
encrypted as is the communication from the ALB to the container application,
ensuring end to end encryption.

Furthermore the application load balancer is also configured to perform
authentication from an OpenID Connect (OIDC) compliant IDP. This eliminates the
need for multiple messages to be passed when using either SAML or OAuth. Upon
authentication, the ALB attaches three fields to the header of the http request
\texttt{\DUrole{code}{x-amzn-oidc-accesstoken}}, \texttt{\DUrole{code}{x-amzn-oidc-identity}} and
\texttt{\DUrole{code}{x-amzn-oidc-data}} which can be used by the end application to confirm
the user's identity and validate the authentication. An example of this process
as implemented in a Jupyter notebook is described below.

For our IDP, we use Okta since it allows us to federate identity services to
additional sign on services. This allows us to onboard collaborators and allow
the collaborators to manage their users.

\paragraph{\textbf{Shared Storage}%
  \label{shared-storage}%
}

In order to facilitate persistence across containers and also collaboration, ECS orchestrates containers on EC2 instances instead of AWS's Fargate product (Fargate facilitates containers in a serverless fashion but does not provide a host to mount an ObjectiveFS file system). Persistent storage can be mounted on the underlying EC2 instances. Individual containers can access the persistent storage by bind mounting the persistent storage. To meet security compliance of encryption at rest, the persistent storage should be encrypted. We elected to use the third party ObjectiveFS for cost reasons though native AWS resources such as elastic file system (EFS) can be used provided that both the file system and the network communications to the file system are encrypted. \DUrole{cite}{efs} ObjectiveFS is a secure file system backed by AWS simple storage service (S3). It should be noted to meet encryption in transit compliance requirements that any network attached storage must have network communications encrypted. For example, the base network file system (\DUroletitlereference{nfs}) protocol is not.

As a specific example with Jupyter notebooks we mount persistent storage as \texttt{\DUrole{code}{/media/home/}}. For a given user say \texttt{\DUrole{code}{user\_a}} we bind mount \texttt{\DUrole{code}{/home/jovyan}} to \texttt{\DUrole{code}{/media/home/user\_a}} so that while in the container the user sees \texttt{\DUrole{code}{/home/jovyan}} the home directory the users files are actually stored in the persistent storage in a \texttt{\DUrole{code}{user\_a}} subdirectory. This configuration has two advantages. Only one persistent volume is needed to support all users' home directories minimizing costs and within the container all users see /home/jovyan thus eliminating the need to build a separate Jupyter container image for each user.

With this configuration, multiple services can use the same home directory. For example, in our R Studio deployment \texttt{\DUrole{code}{/home/rstudio}} is also mapped to \texttt{\DUrole{code}{/media/home/user\_a}}. Furthermore, we also can provide a persistent volume for shared directories. For example, for all users on \texttt{\DUrole{code}{project\_a}} we bind mount \texttt{\DUrole{code}{/home/jovyan/projects/project\_a}} to \texttt{\DUrole{code}{/media/projects/project\_a}} where the persistent volume is mounted to \texttt{\DUrole{code}{/media/projects}}.

\paragraph{\textbf{Resource Summary}%
  \label{resource-summary}%
}

To securely implement the above cloud architecture, each container instance for each user has a set of resources associated with it. First, a task definition is created for each user, this enables customized bind mounts as described above. Additionally, custom environment variables or task commands can also be supplied through the task definition. The task definition can also direct logging the the appropriate AWS CloudWatch stream.

Each user also has a ECS service, ALB listener rule and target group associated with it. This allows the seamless management of connecting a user to the desired container instance.

Finally each service has an AWS IAM role associated with it, this ensures the user has only the access rights to our AWS cloud that are need by the user. Beyond the rights to operate the container task, additional rights might include access to certain S3 storage or certain AWS Secrets Manager. As an example, we use the AWS Secrets Manager to manage user's credentials to various databases and public/private keys.

To simplify management of the per user resources, an AWS CloudFormation template is used to ensure consistency and uniformity among cloud resources whenever a new container instance/user combination is spun up. As an example, our CloudFormation template contains an IAM role, listener rule, target group, task definition, and an ECS service. Each template is then customized to spin up a CloudFormation stack for each user and application combination.

\subsubsection{Container Architecture%
  \label{container-architecture}%
}
\begin{figure}[]\noindent\makebox[\columnwidth][c]{\includegraphics[width=\columnwidth]{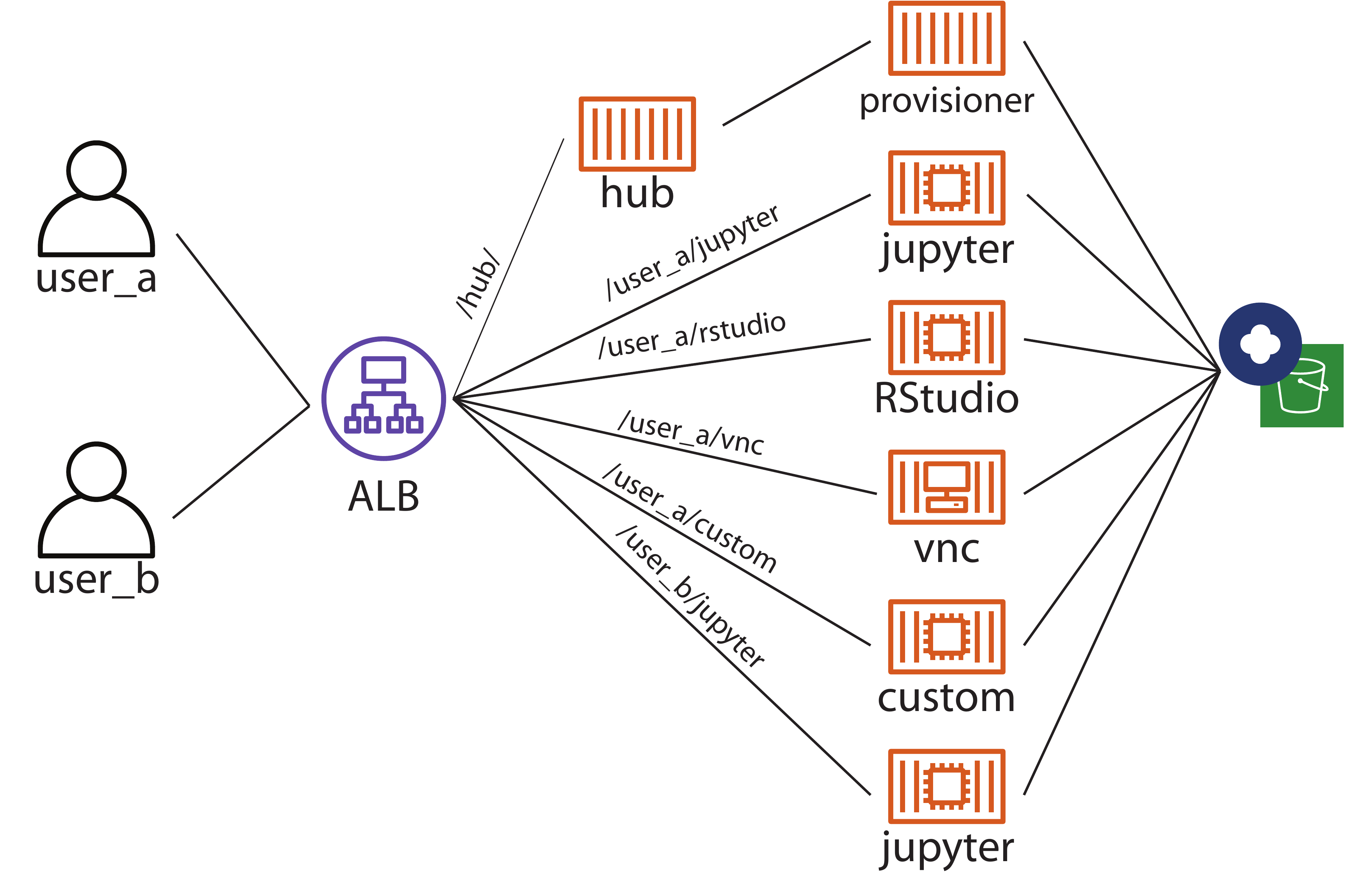}}
\caption{Cloud Architecture. \DUrole{label}{cloudfig}}
\end{figure}

The architecture in terms of container comprises a persistent hub container, an optional ephemeral provisioner container, and an assortment of semi-persistent application containers such as Jupyter notebook. In an alternative deployment, AWS Lambda functions can be functionally substituted for the hub container, but for the sake of simplicity only the container version of the hub is described.

The application containers are described as semi-persistent as they can be started on demand and culled when one or more inactivity criteria has been reached. This can be achieved by updating the associated service to have a desired count of \texttt{\DUrole{code}{1}} to start or a desired count of \texttt{\DUrole{code}{0}} to cull.

We adopted a url path routing convention to access each application such as \texttt{domain.com/user\_id/application}

\paragraph{\textbf{Container Management}%
  \label{container-management}%
}

The heart of the system is the hub container. To facilitate ALB authentication, two listener rules are provided. One rule allows anyone to connect, so that the hub can present a login page (with single sign on and IDP this looks like a single login button). The login action redirects the browser to a url which forces authentication via the ALB. Though this step is not necessary, it provides a cue that makes for a smoother user experience.

Since the hub container may be given privileges to set IAM roles for the application services, the role under which the hub service runs can have a boundary policy attached to it \DUrole{cite}{boundary}. This ensures that any role created by the hub service is constrained to include the boundary policy. This prevents the hub from being able to create an arbitrary role should the container become compromised.

The provisioner container is an ephemeral task which is run with the persistent storage mounted. The provisioner can create a home directory for a user the first time the user logs in and provision the directory with any necessary files. While the functionality of the provisioner container could be incorporated in the hub container. Separation allows the provisioner to run with minimal cloud privileges (IAM role) and allows the hub to have no access to the shared home directory, so in the event the hub container is compromised the user's file system is not exposed. Also, with separation the hub does not have to have access to the file system so it can be refactored and deployed as a Lambda function. Furthermore the provisioner container runs very briefly further limiting the vulnerability window.

Once authenticated, the user can elect to connect to an application container. This can occur under three circumstances: the user's application container is still running, the user's application container has been culled, or the user has never started the application before. If the container is still running, the user is immediately redirected to the container. If the container has been culled, the service is updated to a desired count of \texttt{\DUrole{code}{1}}. If the application has never been started by the user, resources to spin up the service are created such as by creating a CloudFormation stack.

Additionally, an option to \textquotedbl{}decommission\textquotedbl{} an application can be presented where the CloudFormation stack can be deleted.

\paragraph{\textbf{Culling}%
  \label{culling}%
}

The best practice for culling an application is to have the application upon
exiting, set the desired count to \texttt{\DUrole{code}{0}} of its corresponding service.

For the example of Jupyter, the start up scripts for both Jupyter notebook and
Jupyter lab contains the following snippet with \texttt{\DUrole{code}{main}} imported from
different places:\vspace{1mm}
\begin{Verbatim}[commandchars=\\\{\},fontsize=\footnotesize]
\PY{k}{if} \PY{n+nv+vm}{\PYZus{}\PYZus{}name\PYZus{}\PYZus{}} \PY{o}{==} \PY{l+s+s1}{\PYZsq{}}\PY{l+s+s1}{\PYZus{}\PYZus{}main\PYZus{}\PYZus{}}\PY{l+s+s1}{\PYZsq{}}\PY{p}{:}
   \PY{n}{sys}\PY{o}{.}\PY{n}{argv}\PY{p}{[}\PY{l+m+mi}{0}\PY{p}{]} \PY{o}{=} \PY{n}{re}\PY{o}{.}\PY{n}{sub}\PY{p}{(}\PY{l+s+sa}{r}\PY{l+s+s1}{\PYZsq{}}\PY{l+s+s1}{(\PYZhy{}script}\PY{l+s+s1}{\PYZbs{}}\PY{l+s+s1}{.pyw?|}\PY{l+s+s1}{\PYZbs{}}\PY{l+s+s1}{.exe)?\PYZdl{}}\PY{l+s+s1}{\PYZsq{}}\PY{p}{,}
                        \PY{l+s+s1}{\PYZsq{}}\PY{l+s+s1}{\PYZsq{}}\PY{p}{,} \PY{n}{sys}\PY{o}{.}\PY{n}{argv}\PY{p}{[}\PY{l+m+mi}{0}\PY{p}{]}\PY{p}{)}
   \PY{n}{sys}\PY{o}{.}\PY{n}{exit}\PY{p}{(}\PY{n}{main}\PY{p}{(}\PY{p}{)}\PY{p}{)}
\end{Verbatim}
\vspace{1mm}
Rather than just exiting after \texttt{\DUrole{code}{main}} completes, a modified start up script updates the desired count of the corresponding service to \texttt{\DUrole{code}{0}}. Since \texttt{\DUrole{code}{boto3}} essentially wraps API calls to AWS, a delay before termination is needed to ensure the update API call is received before terminating the task. Failure to change the desired count will only result in the service restarting the container upon termination.\vspace{1mm}
\begin{Verbatim}[commandchars=\\\{\},fontsize=\footnotesize]
\PY{k}{if} \PY{n+nv+vm}{\PYZus{}\PYZus{}name\PYZus{}\PYZus{}} \PY{o}{==} \PY{l+s+s1}{\PYZsq{}}\PY{l+s+s1}{\PYZus{}\PYZus{}main\PYZus{}\PYZus{}}\PY{l+s+s1}{\PYZsq{}}\PY{p}{:}
    \PY{n}{sys}\PY{o}{.}\PY{n}{argv}\PY{p}{[}\PY{l+m+mi}{0}\PY{p}{]} \PY{o}{=} \PY{n}{re}\PY{o}{.}\PY{n}{sub}\PY{p}{(}\PY{l+s+sa}{r}\PY{l+s+s1}{\PYZsq{}}\PY{l+s+s1}{(\PYZhy{}script}\PY{l+s+s1}{\PYZbs{}}\PY{l+s+s1}{.pyw?|}\PY{l+s+s1}{\PYZbs{}}\PY{l+s+s1}{.exe)?\PYZdl{}}\PY{l+s+s1}{\PYZsq{}}\PY{p}{,}
                         \PY{l+s+s1}{\PYZsq{}}\PY{l+s+s1}{\PYZsq{}}\PY{p}{,} \PY{n}{sys}\PY{o}{.}\PY{n}{argv}\PY{p}{[}\PY{l+m+mi}{0}\PY{p}{]}\PY{p}{)}
    \PY{n}{main}\PY{p}{(}\PY{p}{)}
    \PY{n}{session} \PY{o}{=} \PY{n}{boto3}\PY{o}{.}\PY{n}{Session}\PY{p}{(}\PY{p}{)}
    \PY{n}{ecs} \PY{o}{=} \PY{n}{session}\PY{o}{.}\PY{n}{client}\PY{p}{(}\PY{l+s+s2}{\PYZdq{}}\PY{l+s+s2}{ecs}\PY{l+s+s2}{\PYZdq{}}\PY{p}{,} \PY{n}{region\PYZus{}name}\PY{p}{)}
    \PY{n}{ecs}\PY{o}{.}\PY{n}{update\PYZus{}service}\PY{p}{(}\PY{n}{cluster}\PY{o}{=}\PY{n}{cluster\PYZus{}name}\PY{p}{,}
                       \PY{n}{service}\PY{o}{=}\PY{n}{service\PYZus{}name}\PY{p}{,}
                       \PY{n}{desiredCount}\PY{o}{=}\PY{l+m+mi}{0}\PY{p}{)}
    \PY{c+c1}{\PYZsh{} Sleep for 2 minutes give service time to update}
    \PY{n}{time}\PY{o}{.}\PY{n}{sleep}\PY{p}{(}\PY{l+m+mi}{120}\PY{p}{)}
\end{Verbatim}
\vspace{1mm}
Code to retrieve the \texttt{\DUrole{code}{region\_name}}, \texttt{\DUrole{code}{cluster\_name}}, and \texttt{\DUrole{code}{service\_name}}, are omitted for clarity, but they can be retrieved from environment variables (set in task definition), passed via \texttt{\DUrole{code}{sys.argv}} or even by calls to \texttt{\DUrole{code}{boto3}}. Though the first two options are simpler.

The above modification to the start up scripts ensures that when Jupyter exits the task count is zero. However, in order for this to be meaningful culling parameters in the Jupyter configuration such as \texttt{\DUrole{code}{c.NotebookApp.shutdown\_no\_activity\_timeout}} \texttt{\DUrole{code}{c.MappingKernelManager.cull\_connected}}, \texttt{\DUrole{code}{c.MappingKernelManager.cull\_idle\_timeout}} and \texttt{\DUrole{code}{c.MappingKernelManager.cull\_interval}}, as well as setting a shell timeout (e.g., \texttt{\DUrole{code}{TMOUT}} environment variable are set) in the event a terminal is open.

\subsection{Authentication and Applications%
  \label{authentication-and-applications}%
}

As mentioned above, the bulk of the authentication is performed by the ALB. However, it is important for the individual application to validate a request forwarded by the ALB, for two reasons. Validation prevents potential security vulnerablities due to a misconfiguration in the system or exposes security vulnerabilities during the initial system debugging. Additionally, validation ensures that the identity of the user is what is expected. The ALB ensures that the user has validly authenticated, but it is up to the application to ensure that the correct user has connected.

Validation is achieved through the JWT token presented in the \texttt{\DUrole{code}{x-amzn-oidc-data}} header by the ALB. These JWT tokens are signed by a public key retrievable from AWS ensuring that only the ALB could have signed them. Within the JWT token, the \texttt{\DUrole{code}{kid}} field represents the \emph{key ID} for the public key. To validate, the key ID should be extracted and corresponding public key should be retrieved from AWS. With the public key, the JWT token can then be validated. We use the \texttt{\DUrole{code}{python-jose}} module available on PyPi. The \texttt{\DUrole{code}{sub}} field in the JWT token is the same as the OIDC ID which is also presented in the \texttt{\DUrole{code}{x-amzn-oidc-identity}} field. The application should then verify this is OIDC ID associated with the expected user.

To deploy an application securely in our infrastructure, in addition to validating the authentication, the application container should meet four more requirements. It should have a configurable base url as the ALB will forward requests to the application with the base url prefix. It should communicate to the ALB over HTTPS to ensure end to end encryption. It should provide a url to respond to pings sent by the ALB for health checks. It should validate that the mounted home container belongs to the user.

The solution to the last requirement is for our provisioner to write an \texttt{.id} file in the user's home directory containing the user's ID. This file is written by \texttt{root} and is only readable. The application upon startup or authenticaation can verify that the user has the correct home directory mounted. This requirement is a safeguard against misconfiguration and can be omitted if one is confident that the system is not misconfigured.

\subsubsection{Jupyter%
  \label{jupyter}%
}

Unfortunately, unlike JupyterHub, Jupyter notebook/lab do not come with a pluggable authentication module. In order to implement validation, the source file \texttt{\DUrole{code}{login.py}} must be modified. This file is usually located in the \texttt{\DUrole{code}{notebook/auth/}} directory in your \texttt{\DUrole{code}{site-packages}} or \texttt{dist-packages} directory. Since Jupyter notebook and JupyterLab are not truly separate applications (in fact they are interchangeable using the path \texttt{/tree} or \texttt{/lab}), the same \texttt{login.py} file facilitates authentication for both. If you build using a standard docker image such as \texttt{\DUrole{code}{jupyter/base-notebook}} or any of its derivative notebooks, this directory would be \texttt{\DUrole{code}{/opt/conda/lib/python3.x/site-packages}} directory. Please note that the specific python version may vary dependent on which version of the docker container is used and whether subsequent additional install modules might force a rollback of python versions.

The specific modification to the \texttt{\DUrole{code}{login.py}} file involves replacing two methods, the \texttt{\DUrole{code}{get}} method and the \texttt{\DUrole{code}{get\_user\_token}} class method of the \texttt{\DUrole{code}{LoginHandler}} class.

Unaltered, the method \texttt{\DUrole{code}{get}} determines whether the \texttt{\DUrole{code}{current\_user}} is set indicating the user has been logged in. If not authenticated, the function presents a login page. Our modification simply adds an additional check that if \texttt{\DUrole{code}{current\_user}} is not set, we validate the JWT token in header to determine additionally whether the user is authenticated. It should also be noted that the function is also decorated as a coroutine to make the function asynchronous as the verification may require network access to retrieve a public key.\vspace{1mm}
\begin{Verbatim}[commandchars=\\\{\},fontsize=\footnotesize]
\PY{n+nd}{@tornado}\PY{o}{.}\PY{n}{gen}\PY{o}{.}\PY{n}{coroutine}
\PY{k}{def} \PY{n+nf}{get}\PY{p}{(}\PY{n+nb+bp}{self}\PY{p}{)}\PY{p}{:}
    \PY{n}{authenticated} \PY{o}{=} \PY{k+kc}{False}
    \PY{k}{if} \PY{n+nb+bp}{self}\PY{o}{.}\PY{n}{current\PYZus{}user}\PY{p}{:}
        \PY{n}{authenticated} \PY{o}{=} \PY{k+kc}{True}
    \PY{k}{else}\PY{p}{:}
        \PY{k}{if} \PY{n+nb+bp}{self}\PY{o}{.}\PY{n}{verify\PYZus{}jwt}\PY{p}{(}\PY{p}{)}\PY{p}{:}
            \PY{n}{authenticated}\PY{o}{=}\PY{k+kc}{True}
    \PY{k}{if} \PY{n}{authenticated}\PY{p}{:}
        \PY{n}{next\PYZus{}url} \PY{o}{=} \PY{n+nb+bp}{self}\PY{o}{.}\PY{n}{get\PYZus{}argument}\PY{p}{(}\PY{l+s+s1}{\PYZsq{}}\PY{l+s+s1}{next}\PY{l+s+s1}{\PYZsq{}}\PY{p}{,}
            \PY{n}{default}\PY{o}{=}\PY{n+nb+bp}{self}\PY{o}{.}\PY{n}{base\PYZus{}url}\PY{p}{)}
        \PY{n+nb+bp}{self}\PY{o}{.}\PY{n}{\PYZus{}redirect\PYZus{}safe}\PY{p}{(}\PY{n}{next\PYZus{}url}\PY{p}{)}
    \PY{k}{else}\PY{p}{:}
        \PY{n+nb+bp}{self}\PY{o}{.}\PY{n}{\PYZus{}render}\PY{p}{(}\PY{p}{)}
\end{Verbatim}
\vspace{1mm}
The other method to be replaced is the \texttt{\DUrole{code}{get\_user\_token}}. Unaltered, the method returns the authorization token used as part of a notebook/lab minimal authentication scheme. This token is normally supplied as a query string in the URL or through the login page. We bypass this mechanism altogether. Instead, we examine the request header for a JWT token supplied by AWS and validate it. If it is successful we provide a token. As far as the rest of the notebook code the value of the token is not used so we supply a random string. Our version of \texttt{\DUrole{code}{get\_user\_token}} uses a local cache to store retrieved public keys and previously the previously decoded user ID.\vspace{1mm}
\begin{Verbatim}[commandchars=\\\{\},fontsize=\footnotesize]
\PY{n+nd}{@classmethod}
\PY{k}{def} \PY{n+nf}{get\PYZus{}user\PYZus{}token}\PY{p}{(}\PY{n+nb+bp}{cls}\PY{p}{,} \PY{n}{handler}\PY{p}{)}\PY{p}{:}
    \PY{l+s+sd}{\PYZdq{}\PYZdq{}\PYZdq{}Identify the user based on}
\PY{l+s+sd}{       Authorization header}

\PY{l+s+sd}{    Returns:}
\PY{l+s+sd}{    \PYZhy{} uuid if authenticated}
\PY{l+s+sd}{    \PYZhy{} None if not}
\PY{l+s+sd}{    \PYZdq{}\PYZdq{}\PYZdq{}}

    \PY{n}{authenticated} \PY{o}{=} \PY{k+kc}{False}
    \PY{k}{if} \PY{n+nb+bp}{cls}\PY{o}{.}\PY{n}{verify\PYZus{}oidc}\PY{p}{(}\PY{n}{handler}\PY{p}{)}\PY{p}{:}
        \PY{n}{authenticated} \PY{o}{=} \PY{k+kc}{True}
    \PY{k}{else}\PY{p}{:}
        \PY{n}{oidc\PYZus{}jwt} \PY{o}{=} \PY{n}{handler}\PY{o}{.}\PY{n}{request}\PY{o}{.}\PY{n}{headers}\PYZbs{}
            \PY{o}{.}\PY{n}{get}\PY{p}{(}\PY{l+s+s1}{\PYZsq{}}\PY{l+s+s1}{x\PYZhy{}amzn\PYZhy{}oidc\PYZhy{}data}\PY{l+s+s1}{\PYZsq{}}\PY{p}{)}
        \PY{k}{if} \PY{n}{oidc\PYZus{}jwt}\PY{p}{:}
            \PY{k}{try}\PY{p}{:}
                \PY{n}{header} \PY{o}{=} \PY{n}{jwt}\PY{o}{.}\PY{n}{get\PYZus{}unverified\PYZus{}headers}\PY{p}{(} \PYZbs{}
                    \PY{n}{oidc\PYZus{}jwt}\PY{p}{)}
            \PY{k}{except} \PY{n}{JOSEError}\PY{p}{:}
                \PY{k}{return} \PY{k+kc}{None}
            \PY{n}{kid} \PY{o}{=} \PY{n}{header}\PY{o}{.}\PY{n}{get}\PY{p}{(}\PY{l+s+s1}{\PYZsq{}}\PY{l+s+s1}{kid}\PY{l+s+s1}{\PYZsq{}}\PY{p}{)}
            \PY{k}{if} \PY{n}{kid} \PY{o+ow}{and} \PY{n}{kid} \PY{o}{==} \PY{n}{user\PYZus{}cache}\PY{o}{.}\PY{n}{get}\PY{p}{(}\PY{l+s+s1}{\PYZsq{}}\PY{l+s+s1}{kid}\PY{l+s+s1}{\PYZsq{}}\PY{p}{)} \PYZbs{}
                \PY{o+ow}{and} \PY{n}{user\PYZus{}cache}\PY{o}{.}\PY{n}{get}\PY{p}{(}\PY{l+s+s1}{\PYZsq{}}\PY{l+s+s1}{pk}\PY{l+s+s1}{\PYZsq{}}\PY{p}{)}\PY{p}{:}
                \PY{k}{try}\PY{p}{:}
                    \PY{n}{token} \PY{o}{=} \PY{n}{jwt}\PY{o}{.}\PY{n}{decode}\PY{p}{(}\PY{n}{oidc\PYZus{}jwt}\PY{p}{,}
                                 \PY{n}{user\PYZus{}cache}\PY{p}{[}\PY{l+s+s1}{\PYZsq{}}\PY{l+s+s1}{pk}\PY{l+s+s1}{\PYZsq{}}\PY{p}{]}\PY{p}{)}
                \PY{k}{except} \PY{n}{JOSEError}\PY{p}{:}
                    \PY{k}{return} \PY{k+kc}{None}
                \PY{n}{oidc\PYZus{}id} \PY{o}{=} \PY{n}{handler}\PY{o}{.}\PY{n}{request}\PY{o}{.}\PY{n}{headers}\PYZbs{}
                   \PY{o}{.}\PY{n}{get}\PY{p}{(}\PY{l+s+s1}{\PYZsq{}}\PY{l+s+s1}{x\PYZhy{}amzn\PYZhy{}oidc\PYZhy{}identity}\PY{l+s+s1}{\PYZsq{}}\PY{p}{)}
                \PY{k}{if} \PY{n}{token}\PY{p}{[}\PY{l+s+s1}{\PYZsq{}}\PY{l+s+s1}{sub}\PY{l+s+s1}{\PYZsq{}}\PY{p}{]} \PY{o}{==} \PY{n}{oidc\PYZus{}id}\PY{p}{:}
                    \PY{n}{authenticated} \PY{o}{=} \PY{k+kc}{True}
                    \PY{n}{user\PYZus{}cache}\PY{p}{[}\PY{l+s+s1}{\PYZsq{}}\PY{l+s+s1}{jwt}\PY{l+s+s1}{\PYZsq{}}\PY{p}{]} \PY{o}{=} \PY{n}{oidc\PYZus{}jwt}
                    \PY{n}{user\PYZus{}cache}\PY{p}{[}\PY{l+s+s1}{\PYZsq{}}\PY{l+s+s1}{user\PYZus{}id}\PY{l+s+s1}{\PYZsq{}}\PY{p}{]} \PY{o}{=} \PY{n}{oidc\PYZus{}id}
    \PY{k}{if} \PY{n}{authenticated}\PY{p}{:}
        \PY{k}{return} \PY{n}{uuid}\PY{o}{.}\PY{n}{uuid4}\PY{p}{(}\PY{p}{)}\PY{o}{.}\PY{n}{hex}
    \PY{k}{else}\PY{p}{:}
        \PY{k}{return} \PY{k+kc}{None}
\end{Verbatim}
\vspace{1mm}
In addition to the two modified methods, we supply two helper methods \texttt{\DUrole{code}{verify\_jwt}} for \texttt{\DUrole{code}{get}} and \texttt{\DUrole{code}{verify\_oidc}} for \texttt{\DUrole{code}{get\_user\_token}}. They perform the token validation and cache management. Additional code which can read identifiers in persistent volumes and verify they match the user who is authenticated can also be added to ensure two authenticated users don't have access to the other's containers.\vspace{1mm}
\begin{Verbatim}[commandchars=\\\{\},fontsize=\footnotesize]
\PY{k}{def} \PY{n+nf}{verify\PYZus{}jwt}\PY{p}{(}\PY{n+nb+bp}{self}\PY{p}{)}\PY{p}{:}
    \PY{k}{global} \PY{n}{user\PYZus{}cache}
    \PY{n}{oidc\PYZus{}id} \PY{o}{=} \PY{n+nb+bp}{self}\PY{o}{.}\PY{n}{request}\PY{o}{.}\PY{n}{headers}\PYZbs{}
                \PY{o}{.}\PY{n}{get}\PY{p}{(}\PY{l+s+s1}{\PYZsq{}}\PY{l+s+s1}{x\PYZhy{}amzn\PYZhy{}oidc\PYZhy{}identity}\PY{l+s+s1}{\PYZsq{}}\PY{p}{)}
    \PY{n}{oidc\PYZus{}jwt} \PY{o}{=} \PY{n+nb+bp}{self}\PY{o}{.}\PY{n}{request}\PY{o}{.}\PY{n}{headers}\PYZbs{}
                \PY{o}{.}\PY{n}{get}\PY{p}{(}\PY{l+s+s1}{\PYZsq{}}\PY{l+s+s1}{x\PYZhy{}amzn\PYZhy{}oidc\PYZhy{}data}\PY{l+s+s1}{\PYZsq{}}\PY{p}{)}

    \PY{k}{if} \PY{o+ow}{not} \PY{n}{oidc\PYZus{}jwt}\PY{p}{:}
        \PY{n+nb+bp}{self}\PY{o}{.}\PY{n}{log}\PY{o}{.}\PY{n}{warning}\PY{p}{(}\PY{l+s+s2}{\PYZdq{}}\PY{l+s+s2}{No JWT Token in Header}\PY{l+s+s2}{\PYZdq{}}\PY{p}{)}
        \PY{k}{return} \PY{k+kc}{False}

    \PY{k}{if} \PY{p}{(}\PY{n}{user\PYZus{}cache}\PY{o}{.}\PY{n}{get}\PY{p}{(}\PY{l+s+s1}{\PYZsq{}}\PY{l+s+s1}{user\PYZus{}id}\PY{l+s+s1}{\PYZsq{}}\PY{p}{)} \PY{o}{==} \PY{n}{oidc\PYZus{}id} \PY{o+ow}{and} \PYZbs{}
        \PY{n}{user\PYZus{}cache}\PY{o}{.}\PY{n}{get}\PY{p}{(}\PY{l+s+s1}{\PYZsq{}}\PY{l+s+s1}{jwt}\PY{l+s+s1}{\PYZsq{}}\PY{p}{)} \PY{o}{==} \PY{n}{oidc\PYZus{}jwt}\PY{p}{)}\PY{p}{:}
        \PY{k}{return} \PY{k+kc}{True}

    \PY{k}{try}\PY{p}{:}
        \PY{n}{header} \PY{o}{=} \PY{n}{jwt}\PY{o}{.}\PY{n}{get\PYZus{}unverified\PYZus{}headers}\PY{p}{(}\PY{n}{oidc\PYZus{}jwt}\PY{p}{)}
    \PY{k}{except} \PY{n}{JOSEError} \PY{k}{as} \PY{n}{e}\PY{p}{:}
        \PY{n+nb+bp}{self}\PY{o}{.}\PY{n}{log}\PY{o}{.}\PY{n}{error}\PY{p}{(}\PY{l+s+s2}{\PYZdq{}}\PY{l+s+s2}{JWT failed to decode: }\PY{l+s+si}{\PYZob{}\PYZcb{}}\PY{l+s+s2}{\PYZdq{}}\PYZbs{}
                   \PY{o}{.}\PY{n}{format}\PY{p}{(}\PY{n}{e}\PY{p}{)}\PY{p}{)}
        \PY{k}{return} \PY{k+kc}{False}

    \PY{n}{kid} \PY{o}{=} \PY{n}{header}\PY{o}{.}\PY{n}{get}\PY{p}{(}\PY{l+s+s1}{\PYZsq{}}\PY{l+s+s1}{kid}\PY{l+s+s1}{\PYZsq{}}\PY{p}{)}
    \PY{k}{if} \PY{o+ow}{not} \PY{n}{kid}\PY{p}{:}
        \PY{n+nb+bp}{self}\PY{o}{.}\PY{n}{log}\PY{o}{.}\PY{n}{error}\PY{p}{(}\PY{l+s+s2}{\PYZdq{}}\PY{l+s+s2}{No Key ID in JWT token}\PY{l+s+s2}{\PYZdq{}}\PY{p}{)}
        \PY{k}{return} \PY{k+kc}{False}

    \PY{k}{if} \PY{n}{kid} \PY{o}{!=} \PY{n}{user\PYZus{}cache}\PY{o}{.}\PY{n}{get}\PY{p}{(}\PY{l+s+s1}{\PYZsq{}}\PY{l+s+s1}{kid}\PY{l+s+s1}{\PYZsq{}}\PY{p}{)}\PY{p}{:}
        \PY{k}{if} \PY{l+s+s1}{\PYZsq{}}\PY{l+s+s1}{pk}\PY{l+s+s1}{\PYZsq{}} \PY{o+ow}{in} \PY{n}{user\PYZus{}cache}\PY{p}{:}
            \PY{k}{del} \PY{n}{user\PYZus{}cache}\PY{p}{[}\PY{l+s+s1}{\PYZsq{}}\PY{l+s+s1}{pk}\PY{l+s+s1}{\PYZsq{}}\PY{p}{]}

    \PY{k}{if} \PY{o+ow}{not} \PY{l+s+s1}{\PYZsq{}}\PY{l+s+s1}{pk}\PY{l+s+s1}{\PYZsq{}} \PY{o+ow}{in} \PY{n}{user\PYZus{}cache}\PY{p}{:}
        \PY{k}{try}\PY{p}{:}
            \PY{n}{r} \PY{o}{=} \PY{n}{requests}\PY{o}{.}\PY{n}{get}\PY{p}{(}\PY{n}{PK\PYZus{}SERVER} \PY{o}{+} \PY{n}{kid}\PY{p}{)}
            \PY{c+c1}{\PYZsh{} TODO treat return code}
            \PY{n}{user\PYZus{}cache}\PY{p}{[}\PY{l+s+s1}{\PYZsq{}}\PY{l+s+s1}{pk}\PY{l+s+s1}{\PYZsq{}}\PY{p}{]} \PY{o}{=} \PY{n}{r}\PY{o}{.}\PY{n}{text}
            \PY{n}{user\PYZus{}cache}\PY{p}{[}\PY{l+s+s1}{\PYZsq{}}\PY{l+s+s1}{kid}\PY{l+s+s1}{\PYZsq{}}\PY{p}{]} \PY{o}{=} \PY{n}{kid}
        \PY{k}{except} \PY{n}{requests}\PY{o}{.}\PY{n}{RequestException} \PY{k}{as} \PY{n}{e}\PY{p}{:}
            \PY{n+nb+bp}{self}\PY{o}{.}\PY{n}{log}\PY{o}{.}\PY{n}{error}\PY{p}{(}\PY{l+s+s2}{\PYZdq{}}\PY{l+s+s2}{Requests Error: }\PY{l+s+si}{\PYZob{}\PYZcb{}}\PY{l+s+s2}{\PYZdq{}}\PYZbs{}
                       \PY{o}{.}\PY{n}{format}\PY{p}{(}\PY{n}{e}\PY{p}{)}\PY{p}{)}
            \PY{k}{return} \PY{k+kc}{False}

    \PY{k}{try}\PY{p}{:}
        \PY{n}{token} \PY{o}{=} \PY{n}{jwt}\PY{o}{.}\PY{n}{decode}\PY{p}{(}\PY{n}{oidc\PYZus{}jwt}\PY{p}{,}
                           \PY{n}{user\PYZus{}cache}\PY{p}{[}\PY{l+s+s1}{\PYZsq{}}\PY{l+s+s1}{pk}\PY{l+s+s1}{\PYZsq{}}\PY{p}{]}\PY{p}{)}
    \PY{k}{except} \PY{n}{JOSEError} \PY{k}{as} \PY{n}{e}\PY{p}{:}
        \PY{n+nb+bp}{self}\PY{o}{.}\PY{n}{log}\PY{o}{.}\PY{n}{info}\PY{p}{(}\PY{l+s+s2}{\PYZdq{}}\PY{l+s+s2}{JWT failed to validate: }\PY{l+s+si}{\PYZob{}\PYZcb{}}\PY{l+s+s2}{\PYZdq{}}\PYZbs{}
                   \PY{o}{.}\PY{n}{format}\PY{p}{(}\PY{n}{e}\PY{p}{)}\PY{p}{)}
        \PY{k}{return} \PY{k+kc}{False}

    \PY{k}{if} \PY{n}{token}\PY{p}{[}\PY{l+s+s1}{\PYZsq{}}\PY{l+s+s1}{sub}\PY{l+s+s1}{\PYZsq{}}\PY{p}{]} \PY{o}{!=} \PY{n}{oidc\PYZus{}id}\PY{p}{:}
        \PY{n+nb+bp}{self}\PY{o}{.}\PY{n}{log}\PY{o}{.}\PY{n}{error}\PY{p}{(}\PY{l+s+s2}{\PYZdq{}}\PY{l+s+s2}{User ID in token doesn}\PY{l+s+s2}{\PYZsq{}}\PY{l+s+s2}{t }\PY{l+s+s2}{\PYZdq{}}
                       \PY{l+s+s2}{\PYZdq{}}\PY{l+s+s2}{match user ID in header}\PY{l+s+s2}{\PYZdq{}}\PY{p}{)}
        \PY{k}{return} \PY{k+kc}{False}

    \PY{n}{user\PYZus{}cache}\PY{p}{[}\PY{l+s+s1}{\PYZsq{}}\PY{l+s+s1}{user\PYZus{}id}\PY{l+s+s1}{\PYZsq{}}\PY{p}{]} \PY{o}{=} \PY{n}{oidc\PYZus{}id}
    \PY{n}{user\PYZus{}cache}\PY{p}{[}\PY{l+s+s1}{\PYZsq{}}\PY{l+s+s1}{jwt}\PY{l+s+s1}{\PYZsq{}}\PY{p}{]} \PY{o}{=} \PY{n}{oidc\PYZus{}jwt}

\PY{n+nd}{@classmethod}
\PY{k}{def} \PY{n+nf}{verify\PYZus{}oidc}\PY{p}{(}\PY{n+nb+bp}{cls}\PY{p}{,} \PY{n}{handler}\PY{p}{)}\PY{p}{:}
    \PY{k}{global} \PY{n}{user\PYZus{}cache}
    \PY{n}{oidc\PYZus{}id} \PY{o}{=} \PY{n}{handler}\PY{o}{.}\PY{n}{request}\PY{o}{.}\PY{n}{headers}\PYZbs{}
                \PY{o}{.}\PY{n}{get}\PY{p}{(}\PY{l+s+s1}{\PYZsq{}}\PY{l+s+s1}{x\PYZhy{}amzn\PYZhy{}oidc\PYZhy{}identity}\PY{l+s+s1}{\PYZsq{}}\PY{p}{)}
    \PY{n}{oidc\PYZus{}jwt} \PY{o}{=} \PY{n}{handler}\PY{o}{.}\PY{n}{request}\PY{o}{.}\PY{n}{headers}\PYZbs{}
                \PY{o}{.}\PY{n}{get}\PY{p}{(}\PY{l+s+s1}{\PYZsq{}}\PY{l+s+s1}{x\PYZhy{}amzn\PYZhy{}oidc\PYZhy{}data}\PY{l+s+s1}{\PYZsq{}}\PY{p}{)}

    \PY{k}{if} \PY{o+ow}{not} \PY{n}{oidc\PYZus{}id} \PY{o+ow}{or} \PY{o+ow}{not} \PY{n}{oidc\PYZus{}jwt}\PY{p}{:}
        \PY{k}{return} \PY{k+kc}{False}
    \PY{k}{if} \PY{n}{oidc\PYZus{}id} \PY{o}{!=} \PY{n}{user\PYZus{}cache}\PY{o}{.}\PY{n}{get}\PY{p}{(}\PY{l+s+s1}{\PYZsq{}}\PY{l+s+s1}{user\PYZus{}id}\PY{l+s+s1}{\PYZsq{}}\PY{p}{)}\PY{p}{:}
        \PY{k}{return} \PY{k+kc}{False}
    \PY{k}{if} \PY{n}{oidc\PYZus{}jwt} \PY{o}{!=} \PY{n}{user\PYZus{}cache}\PY{o}{.}\PY{n}{get}\PY{p}{(}\PY{l+s+s1}{\PYZsq{}}\PY{l+s+s1}{jwt}\PY{l+s+s1}{\PYZsq{}}\PY{p}{)}\PY{p}{:}
        \PY{k}{return} \PY{k+kc}{False}
    \PY{k}{try}\PY{p}{:}
        \PY{n}{header} \PY{o}{=} \PY{n}{jwt}\PY{o}{.}\PY{n}{get\PYZus{}unverified\PYZus{}headers}\PY{p}{(}\PY{n}{oidc\PYZus{}jwt}\PY{p}{)}
    \PY{k}{except} \PY{n}{JOSEError}\PY{p}{:}
        \PY{k}{return} \PY{k+kc}{False}
    \PY{n}{kid} \PY{o}{=} \PY{n}{header}\PY{o}{.}\PY{n}{get}\PY{p}{(}\PY{l+s+s1}{\PYZsq{}}\PY{l+s+s1}{kid}\PY{l+s+s1}{\PYZsq{}}\PY{p}{)}
    \PY{k}{if} \PY{n}{kid} \PY{o}{!=} \PY{n}{user\PYZus{}cache}\PY{o}{.}\PY{n}{get}\PY{p}{(}\PY{l+s+s1}{\PYZsq{}}\PY{l+s+s1}{kid}\PY{l+s+s1}{\PYZsq{}}\PY{p}{)}\PY{p}{:}
        \PY{k}{return} \PY{k+kc}{False}

    \PY{k}{return} \PY{k+kc}{True}
\end{Verbatim}
\vspace{1mm}
To meet the other requirements for Jupyter, the \texttt{\DUrole{code}{base\_url}} configuration needs to be set to ensure that the route is properly interpreted. Furthermore, we use this \texttt{base\_url} as the health check url which responds with a \texttt{302} code. A self-signed certificate is automatically generated when the container starts and that certificate is then used to configure Jupyter to run over HTTPS.

\subsubsection{RStudio%
  \label{rstudio}%
}
\begin{figure}[]\noindent\makebox[\columnwidth][c]{\includegraphics[width=\columnwidth]{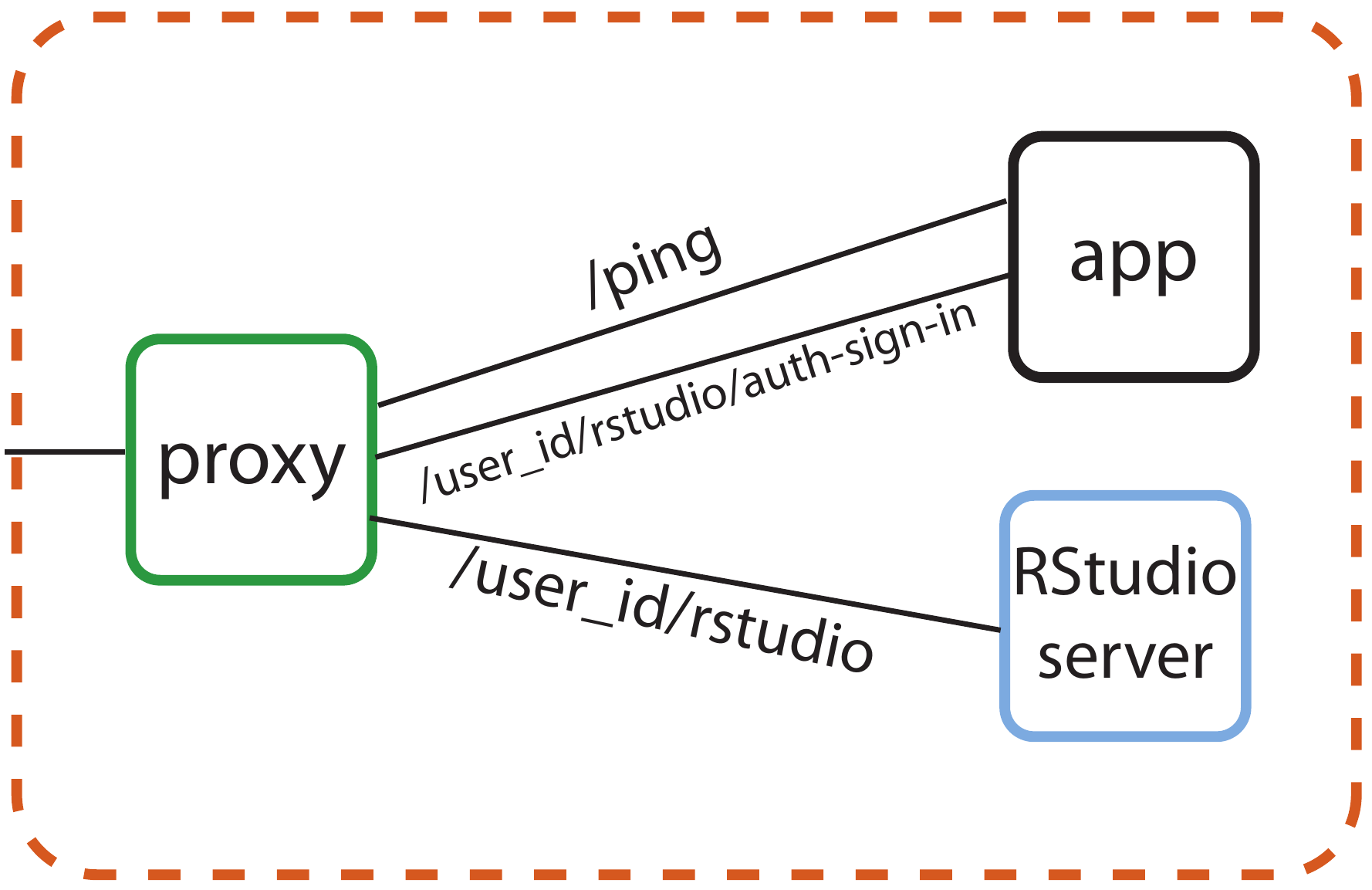}}
\caption{Inside the RStudio Container \DUrole{label}{rstudio}}
\end{figure}

Our implementation of RStudio Server on the same cloud platform is non-invasive to the code base, but more complicated architecturally. Since RStudio does not have a way to set the base URL of the application, a proxy is required to rewrite the HTTPS request paths. We use an \texttt{\DUrole{code}{nginx}} proxy to rewrite requests to RStudio Server using the \texttt{\DUrole{code}{proxy\_redirect}} directive.

Figure \DUrole{ref}{rstudio} shows the application structure within the RStudio container. A proxy communicates with the ALB and routes some requests to a custom app used for authentication and handling the health checks and others to the RStudio server. Since communications between the proxy, app and RStudio server are all within the container and not exposed, they do not require encryption to satisfy compliance. A self-signed certificate is created upon container startup that enables \texttt{nginx} to communicate over HTTPS to the ALB.

For authentication, RStudio Server maintains authentication session information in a cookie. So with \texttt{\DUrole{code}{nginx}} we capture, the \texttt{\DUrole{code}{auth-sign-in}} URL and redirect it to an lightweight webapp whose sole function is to authenticate the user, set the cookie and redirect the browser to RStudio Server. Since the app is necessary in this configuration, we also configure the app to respond to a \texttt{\DUrole{code}{/ping}} request issued by the ALB target group's health check.

The authentication code is nearly identical to the \texttt{\DUrole{code}{verify\_jwt}} function written above for Jupyter. The cookie consists of three pieces, a user ID (which we retain as the default \texttt{\DUrole{code}{rstudio}} as we retained \texttt{\DUrole{code}{jovyan}} for the Jupyter notebook, to prevent the need to build a separate docker image for each user), the expiry and an HMAC 256 signature, signed with a secret typically stored at \texttt{\DUrole{code}{/var/lib/rstudio-server/secure-cookie-key}} inside the container. The following snippet of code implements this.\vspace{1mm}
\begin{Verbatim}[commandchars=\\\{\},fontsize=\footnotesize]
\PY{k+kn}{from} \PY{n+nn}{urllib}\PY{n+nn}{.}\PY{n+nn}{parse} \PY{k+kn}{import} \PY{n}{quote}
\PY{k+kn}{from} \PY{n+nn}{Crypto}\PY{n+nn}{.}\PY{n+nn}{Hash} \PY{k+kn}{import} \PY{n}{HMAC}
\PY{k+kn}{from} \PY{n+nn}{Crypto}\PY{n+nn}{.}\PY{n+nn}{Hash} \PY{k+kn}{import} \PY{n}{SHA256}
\PY{k+kn}{import} \PY{n+nn}{base64}
\PY{k+kn}{import} \PY{n+nn}{datetime}

\PY{n}{utc} \PY{o}{=} \PY{n}{datetime}\PY{o}{.}\PY{n}{datetime}\PY{o}{.}\PY{n}{utcnow}\PY{p}{(}\PY{p}{)}
\PY{n}{expiry} \PY{o}{=} \PY{n}{utc} \PY{o}{+} \PY{n}{datetime}\PY{o}{.}\PY{n}{timedelta}\PY{p}{(}\PY{n}{days}\PY{p}{)}
\PY{n}{now} \PY{o}{=} \PY{n}{expiry}\PY{o}{.}\PY{n}{strftime}\PY{p}{(}\PY{l+s+s1}{\PYZsq{}}\PY{l+s+si}{\PYZpc{}a}\PY{l+s+s1}{, }\PY{l+s+si}{\PYZpc{}d}\PY{l+s+s1}{ }\PY{l+s+s1}{\PYZpc{}}\PY{l+s+s1}{b }\PY{l+s+s1}{\PYZpc{}}\PY{l+s+s1}{Y }\PY{l+s+s1}{\PYZpc{}}\PY{l+s+s1}{H:}\PY{l+s+s1}{\PYZpc{}}\PY{l+s+s1}{M:}\PY{l+s+s1}{\PYZpc{}}\PY{l+s+s1}{S GMT}\PY{l+s+s1}{\PYZsq{}}\PY{p}{)}
\PY{n}{dig} \PY{o}{=} \PY{n}{base64}\PY{o}{.}\PY{n}{b64encode}\PY{p}{(} \PYZbs{}
        \PY{n}{HMAC}\PY{o}{.}\PY{n}{new}\PY{p}{(}\PY{n}{secret}\PY{p}{,}
                 \PY{l+s+s2}{\PYZdq{}}\PY{l+s+si}{\PYZob{}0\PYZcb{}}\PY{l+s+si}{\PYZob{}1\PYZcb{}}\PY{l+s+s2}{\PYZdq{}}\PY{o}{.}\PY{n}{format}\PY{p}{(}\PY{n}{username}\PY{p}{,} \PY{n}{now}\PY{p}{)}\PY{p}{,}
                 \PY{n}{digestmod}\PY{o}{=}\PY{n}{SHA256}\PY{p}{)}\PY{o}{.}\PY{n}{digest}\PY{p}{(}\PY{p}{)}\PY{p}{)}

\PY{n}{cookie} \PY{o}{=} \PY{n}{quote}\PY{p}{(}\PY{l+s+s2}{\PYZdq{}}\PY{l+s+si}{\PYZob{}0\PYZcb{}}\PY{l+s+s2}{|}\PY{l+s+si}{\PYZob{}1\PYZcb{}}\PY{l+s+s2}{|}\PY{l+s+si}{\PYZob{}2\PYZcb{}}\PY{l+s+s2}{\PYZdq{}}\PY{o}{.}\PY{n}{format}\PY{p}{(}\PY{n}{username}\PY{p}{,}
                                    \PY{n}{now}\PY{p}{,}
                                    \PY{n}{dig}\PY{o}{.}\PY{n}{decode}\PY{p}{(}\PY{p}{)}\PY{p}{)}\PY{p}{,}
               \PY{l+s+s1}{\PYZsq{}}\PY{l+s+s1}{|}\PY{l+s+s1}{\PYZsq{}}\PY{p}{)}
\PY{n}{response}\PY{o}{.}\PY{n}{set\PYZus{}cookie}\PY{p}{(}\PY{l+s+s1}{\PYZsq{}}\PY{l+s+s1}{user\PYZhy{}id}\PY{l+s+s1}{\PYZsq{}}\PY{p}{,} \PY{n}{cookie}\PY{p}{)}
\end{Verbatim}
\vspace{1mm}
The \texttt{\DUrole{code}{days}} is the number of days til the cookie expires, and \texttt{\DUrole{code}{username}} is the user name (i.e. \texttt{\DUrole{code}{rstudio}}). In the above snippet, the cookie is attacked to a Flask response.

\subsubsection{Virtual Network Computing (VNC) Containers%
  \label{virtual-network-computing-vnc-containers}%
}
\begin{figure}[]\noindent\makebox[\columnwidth][c]{\includegraphics[width=\columnwidth]{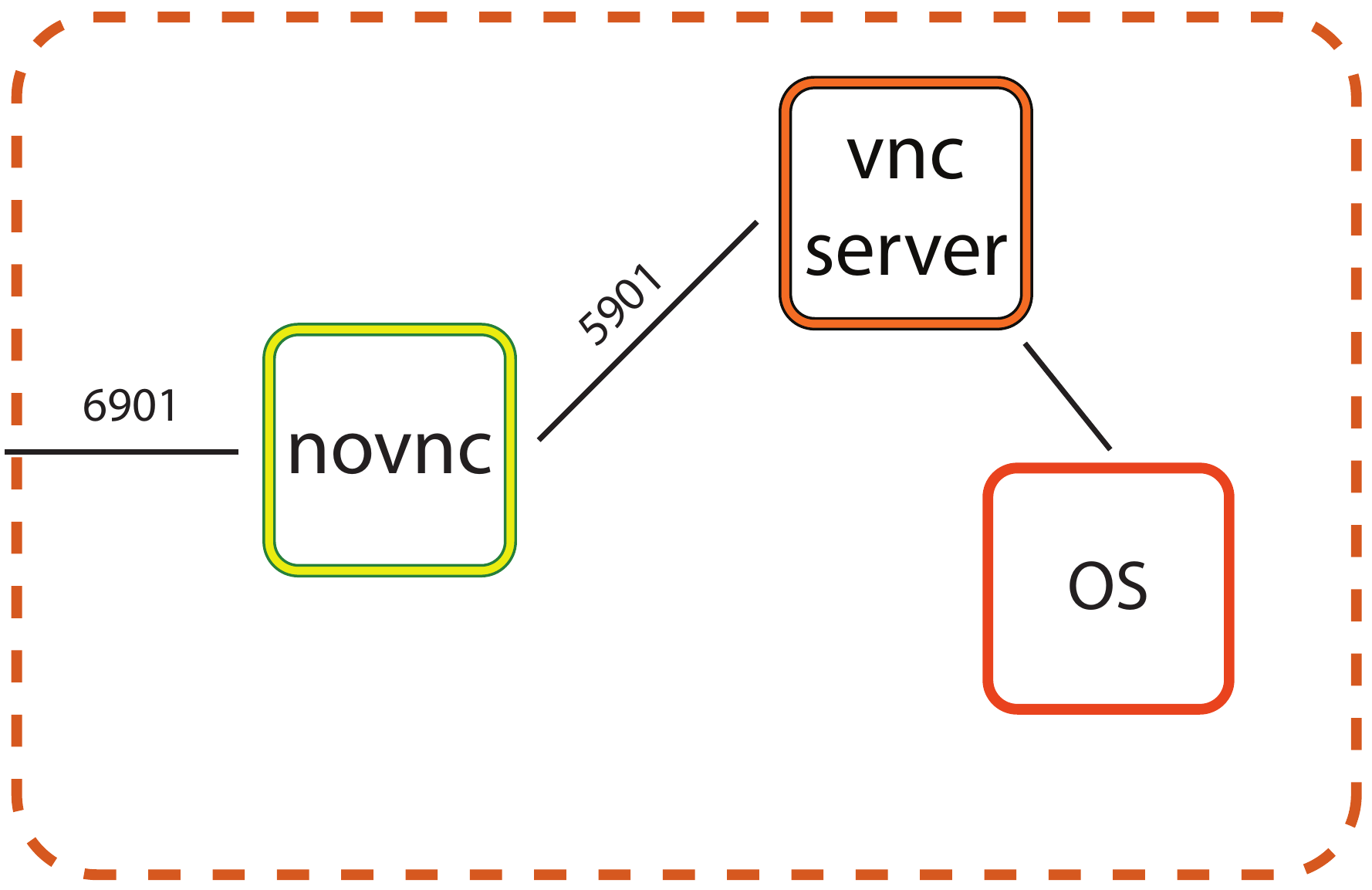}}
\caption{Inside a VNC Container \DUrole{label}{vnc}}
\end{figure}

There are many desktop apps for Linux which may also be useful to deploy via a web application on a cloud cluster such as presented here. The following implementation allows the deployment of such applications such as Orange and Falcon through the use of a web VNC client to a VNC server running in a container.

This is based on the Docker Headless VNC Container project \DUrole{cite}{headless} as a blueprint using the \texttt{\DUrole{code}{xfce4}} window manager. Since it appears that the project has been inactive for over a year we adopt its \texttt{\DUrole{code}{Dockerfile}} as a starting point but do not use the docker images as a building block.

Figure \DUrole{ref}{vnc} shows the application structure within a headless VNC container. NoVNC \DUrole{cite}{novnc} is used as a web to vnc proxy which connects via VNC to a local vnc server which in accordance to the Docker Headless VNC Container project is tigerVNC \DUrole{cite}{tiger}. Through the VNC server graphically oriented operating system commands and applications can be executed. In our container tigerVNC is unchanged and is installed just as it is in the headerless project's \texttt{\DUrole{code}{Dockerfile}}. The noVNC project comprises a \texttt{\DUrole{code}{novnc}} and \texttt{\DUrole{code}{websockify}} component. No changes were made to the \texttt{\DUrole{code}{novnc}} component except to alter the parameters use to start \texttt{\DUrole{code}{websockify}}. Therefore the focus of the customization is on the \texttt{\DUrole{code}{websockify}} component.

Fortunately, \texttt{\DUrole{code}{websockify}} permits authentication plugins. The plugin is a simple class with an \texttt{\DUrole{code}{authenticate}} method which accepts the \texttt{\DUrole{code}{headers}}, \texttt{\DUrole{code}{target\_host}} and \texttt{\DUrole{code}{target\_port}} as parameters. Upon success it returns and on failure it raises an \texttt{\DUrole{code}{AuthenticationError}} exception. Since the body of the code is essentially the same as the \DUroletitlereference{verify\_jwt} method descirbed for Jupyter, the code is not repeated here.

It should be noted that in the container by default the VNC server listens on port 5901 and the novnc client listens on port 6901. It is recommended that only port 6901 be exposed so that only the novnc client can directly communicate with the VNC server as the VNC password in this environment is not well protected. By only exposing port 6901, knowledge of the VNC password can not be exploited to bypass the authentication.

Furthermore, the web server within the \texttt{\DUrole{code}{websockify}} project is located in \texttt{\DUrole{code}{websockifyserver.py}} and is based on \texttt{\DUrole{code}{SimpleHTTPServer}}. It may be desirable to create a custom handler or custom \texttt{\DUrole{code}{do\_GET}} method to handle issues such as providing a base URL, health check URL for the ALB's target group, or to implement templating if desired. A self-signed certificate is generated in a \DUroletitlereference{launch.sh} as \DUroletitlereference{self.pem} which the webserver will automatically detect and run using HTTPS.

Once this base container image is built with those customizations. Applications
such as Orange or Falcon can be added, thus not limiting the cloud system to
web applications.

\subsubsection{Custom Applications%
  \label{custom-applications}%
}

In developing your own bespoke applications, a layer of authentication can be employed. In consideration of developing or adapting your own application, you should provide an unauthenticated URL for the ALB's health check and be equipped to configure the base URL. Authentication can be easily plugged into most web server frameworks.

As a simple example, using flask authentication can be incorporated into a custom \texttt{\DUrole{code}{login\_required}} decorator, so that for any protected URL the request is authenticated before being processed. Once again the decorator could be implemented with code similar to that of \DUroletitlereference{jwt\_verify} described above.

\subsection{Security and Compliance%
  \label{security-and-compliance}%
}

In our cloud architecture, the bulk of the security and compliance is built into the EC2 instances serving as nodes behind the ALB and built into features of the ALB. By keeping most of the security external to the containers, container images need less customization for security purposes making it easier to support a wide variety of container images and container apps.

The preferred method to implement security, compliance, and even maintenance services on an EC2 instance is to install the appropriate software in an Amazon machine image (AMI). By building a customized AMI based off an optimized Amazon ECS reference AMI \DUrole{cite}{aws:ecs} but including the desired additional services installed, an fully equipped EC2 instance can be spun up quickly and features such as autoscaling can easily be applied.

Specifics to security and compliance implementations are described in the following subsections including encryption at rest, access controls, auditing and other agents.

\subsubsection{Encryption at Rest%
  \label{encryption-at-rest}%
}

As previoiusly mentioned, persistent storage and associated file system
protocol are encrypted give both encryption in transit and encryption at rest
for the persistent storage. However, it is also important that the base file
system of the EC2 instances are also encrypted to fully ensure encryption at
rest. There are two important aspects of ensuring encryption at rest for the
base file system. First the attached file system such as elastic block
storage (EBS) must be encrypted. This is accomplished by selecting
encryption when creating the EC2 instance or within a launch configuration.
Fortunately, AWS now offers an account-level option where EBS volumes are
encrypted by default for any EBS volumes created in that given account. We
highly recommend this option as it will mitigate the chances of
misconfiguration.

Furthermore, the AMI used to create EC2 instances must also be encrypted. A common technique for doing so is to build an machine snapshot will all the agents and services desired then encrypt the snapshot. Regardless for what techinque is used. the AMI's should be encrypted to satisfy any requirements for encryption at rest.

\subsubsection{Access Control%
  \label{access-control}%
}

Another security concern is controlling the internet access from the container. The reason is two fold. First, controlling access allows us to prevent users from within a container from accessing potentially malicious websites. Second, should a container become compromised we want to mitigate the compromised container's ability to escalate privileges or pivot to other services within the organization. While AWS through the use of security groups and access control lists provide a coarse ability to regulate what destinations are accessible, we favor more fine grain control.

There are two aspects of this finer grain control, first we use an on-host firewall to control outbound access from the hosted containers. Second we funnel all traffic from each container to a proxy.

For the firewall, we use \texttt{iptables} using the following commands:\vspace{1mm}
\begin{Verbatim}[commandchars=\\\{\},fontsize=\footnotesize]
iptables \PYZhy{}\PYZhy{}insert DOCKER\PYZhy{}USER \PYZhy{}\PYZhy{}in\PYZhy{}interface docker0 \PY{l+s+se}{\PYZbs{}}
   \PYZhy{}o eth0 \PYZhy{}j DROP

iptables \PYZhy{}\PYZhy{}insert DOCKER\PYZhy{}USER \PY{l+s+se}{\PYZbs{}}
   \PYZhy{}\PYZhy{}destination 169.254.169.254 \PYZhy{}\PYZhy{}jump REJECT \PY{l+s+se}{\PYZbs{}}
   \PYZhy{}\PYZhy{}reject\PYZhy{}with icmp\PYZhy{}port\PYZhy{}unreachable

iptables \PYZhy{}t nat \PYZhy{}A PREROUTING \PYZhy{}i docker0 \PY{l+s+se}{\PYZbs{}}
   \PYZhy{}d 172.17.0.1 \PYZhy{}p tcp \PYZhy{}\PYZhy{}dport 8888 \PYZhy{}j RETURN
iptables \PYZhy{}t nat \PYZhy{}A PREROUTING \PYZhy{}i docker0 \PY{l+s+se}{\PYZbs{}}
   \PYZhy{}d 172.17.0.1 \PYZhy{}p tcp \PYZhy{}j DNAT \PYZhy{}\PYZhy{}to\PYZhy{}destination :2
\end{Verbatim}
\vspace{1mm}
The first command blocks all internet traffic coming from the \DUroletitlereference{docker0} interface (where the containers must route through) to the \DUroletitlereference{eth0} interface which is the external interface. The second command (see \DUrole{cite}{costa}) blocks access to the node specific metadata service, which typical contains information about the EC2 instance and credentials for that instance.d. Blocking this prevents a compromised container from accessing the metadata about the EC2 instances blocking a potential escalation in privliges to that of the EC2 node. The third and fourth commands allows the container access to the EC2 instance (which in the docker world is IP address \texttt{172.17.0.1}) only on port 8888, where the proxy is configured to listen. All other access is routed to port 2 which has no active listeners.

On the container side, the environment variables \texttt{http\_proxy} and \texttt{https\_proxy} must be set to forward all http and https request to the EC2 instance at port 8888. In addition the \texttt{no\_proxy} environment variable should be set to allow some traffic not to be forced into the proxy. Of course, \texttt{localhost} (and corresponding IP address \texttt{127.0.0.1}) do not require proxy as the traffic doesn't leave the container. In addition, the metadata IP address \texttt{169.254.169.254} should be allowed out so that the \texttt{iptables} rule regarding the metadata traffic can be enforced. Finally, the IP address \texttt{169.254.169.2} is used by the ECS agent.

Two methods can be used to address the environment variables. Either we can add the environment variables to the task definition when an application service created or it can defined in the container's \DUroletitlereference{Dockerfile} with the following lines:\vspace{1mm}
\begin{Verbatim}[commandchars=\\\{\},fontsize=\footnotesize]
ENV \PY{n+nv}{http\PYZus{}proxy}\PY{o}{=}http://172.17.0.1:8888/
ENV \PY{n+nv}{https\PYZus{}proxy}\PY{o}{=}http://172.17.0.1:8888/
ENV \PY{n+nv}{no\PYZus{}proxy}\PY{o}{=}localhost,127.0.0.1,\PY{l+s+se}{\PYZbs{}}
169.254.169.254,169.254.170.2
\end{Verbatim}
\vspace{1mm}
Because of the \texttt{iptables} rules a misconfiguration that fails to set the proper environment variables results in loss of access and not a vulnerability.

The proxy can then determine whether to route the connection request directly externally or through an external outbound gateway which could include a company firewall so that broad based policies could be applied. For the proxy we selected \texttt{tinyproxy} because it is lightweight and allows gateway credentials to be embedded in the proxy configuration pushing the burden of gateway credentials to the proxy and not the container or application of the container.

\subsubsection{Auditing%
  \label{auditing}%
}

Beyond security reasons, many regulations such as HIPAA require auditing for compliance. Our approach is two fold. We use the ALB logging capabilities to track access to application containers and authentication. We use a logging agent to track potential privlege escalation or other security concerns on the underlying EC2 host.

The ALB provides logging \DUrole{cite}{aws:alb:logging} which will log all access to the application containers to an S3 bucket. Because in our architecture all authentication is performed using the ALB all authentication attempts both successful and more importantly failures are also logged to the bucket. Many third party log management tools are configurable to digest logs stored in this manner including Loggly, Splunk, Sumo Logic.

Another good practice is to set the target S3 bucket in a separate AWS account and only grant privleges to the logging account to write to the bucket but not delete. This ensure that even if a container or the EC2 instance is compromised, the logs can not be tampered with.

To supplement the auditing and monitoring capability one or more logging agents are installed on the EC2 instance. Essentially, this agent transmits logs of interest such as the system log \texttt{syslog} to an external log management system. Through this mechanism behaviours such as privlege escalation (e.g. \texttt{sudo}) are tracked. We use both the native AWS logging agent and a third party logging agent.

With both mechanisms in place, the preferred log management system can be configured to provide alarms when severe incidents occurs and generate reports of incidents as may be required by compliance requirements.

\subsubsection{Other Useful Agents%
  \label{other-useful-agents}%
}

Building a custom AMI image to spin up an EC2 instance to support our ECS cluster affords the opportunity to install additional agents to meet security, compliance and maintenance needs. Our best practices is to the include the following additional agents in the AMI. Some of agents are provided by AWS while some are third party.

\paragraph{\textbf{ECS Agent}%
  \label{ecs-agent}%
}

The AWS ECS agent is required in order for the EC2 instance to serve ECS containers. However, periodically updating the ECS agent is important in that potential vulnerabilites may be fixed and newer agents offer more features to aid in maintenance. Furthermore, proper configuration of features can aid in security as well. For example, the ECS agent can be configure so that the maximum lifetime of an EC2 instance is set. This is particularly useful if the AMIs for the EC2 instances are constantly being updated with security patches etc. The limited lifetime guarantees that the EC2 instances running will not be based on an AMI that is too out of date.

\paragraph{\textbf{Systems Manager Agent}%
  \label{systems-manager-agent}%
}

Another useful AWS Agent that can be employed is the AWS Systems Manager Agent (SSM) \DUrole{cite}{aws:ssm}. The SSM agent allows the ``Systems Manager to update, manage and configure'' the EC2 instances. This agent makes it easier to maintain EC2 instances in a centralized manner. Once again keeping an EC2 instance up to date helps reduce vulnerabilites on the node.

\paragraph{\textbf{Anti-virus}%
  \label{anti-virus}%
}

An antivirus or antimalware agent is also recommended. The antivirus should be
one that is container aware and that the container awareness feature should be
active. This would facilitate pinpointing the specific container that may be
compromised. Container systems such as docker are not complete virtualizations.
Processes that run in a container run as processes in the native host, as such
an antivirus agent inside can monitor processes that occur ``inside a
container''. Container aware antivirus agents makes mitigation in a container
environment easier. In our particular configuration, we use Sophos as the
antivirus but you may have your own preferences.

\paragraph{\textbf{Intrusion Detection}%
  \label{intrusion-detection}%
}

Another useful agent to be deployed on the EC2 instance is an intrusion
detection agent. Like this antivirus agent, an intrusion detection agent that
has container awareness capabilities is desireable and should have the
capability activated. The intrusion detection agent looks for activities that
are anomolous and when high risk activity is detected, it will gather as much
information around the incident as it can. We use ThreatStack for our intrusion
detection.

\subsection{Conclusion%
  \label{conclusion}%
}

Presented here is a secure, collaborative infrastructure for deploying a cloud
computation resources. The primary purpose of our infrastructure is to provide
Jupyter in this environment though due to the preference of some of our users
RStudio and other tools are include. Our Data Science and infrastructure team
is small so building a compliant infrastructure that requires little
maintenance is paramount. Equally important is to safeguard against opening
vulnerabilities due to misconfigurations. By following the suggestions
presented here, misconfigurations err on the side of loss of functionality
rather than introducing vulnerabilites.

The architecture presented here was successful in a recently performed
penetration test. While the recommendations and architecture shown here rely
heavily on AWS resources. No doubt elements and counterparts can be found in
other cloud services such as Google Cloud and Microsoft Azure.

Snippets of code, Dockerfile, commands and other resources presented here and
the corresponding poster are available at West Health's github repository at
\href{https://github.com/Westhealth/scipy2020/cloud_infrastructure/}{https://github.com/Westhealth/scipy2020/cloud\_infrastructure}.
\bibliographystyle{alphaurl}
\bibliography{ourbib}

\end{document}